\documentstyle[preprint,aps]{revtex}
\input epsf.tex
\def\DESepsf(#1 width #2){\epsfxsize=#2 \epsfbox{#1}}
\begin{document}
\preprint{\vbox{\hbox{}}}
\draft
\title{
SU(3) Flavor Symmetry and CP Violating Rate Differences
for Charmless $B\to P V$ Decays}
\author{N. G. Deshpande$^1$, Xiao-Gang He$^2$, and Jian-Qing Shi$^2$}
\address{$^1$Institute of Theoretical Sciences, University of Oregon,
OR 97403, USA\\
$^2$Department of Physics, National Taiwan University, Taipei, Taiwan 10764, R.O.C.
}
\date{February, 2000}
\maketitle
\begin{abstract}
We derive several  relations between
CP violating rate differences $\Delta(B\to PV) = \Gamma( B\to PV) -
\Gamma(\bar B \to \bar P \bar V)$
for charmless $B\to P V$ decays in the Standard Model using SU(3) flavor symmetry.
It is found that
although the relations between branching ratios of $\Delta S = 0$ and $\Delta S = -1$
processes are complicated, there are simple relations independent of hadronic 
models between some of the $\Delta S =0$ and
$\Delta S = -1$ rate differences
 due to the unitarity property of the Kobayashi-Maskawa matrix,
such as $ \Delta(B\to \pi^+ \rho^-) = -\Delta(B\to \pi^+ K^{*-})$,
$\Delta(B\to \pi^- \rho^+) = - \Delta(K^- \rho^+)$. SU(3) breaking effects
are also estimated using factorization approximation.
These relations can be tested at B factories in the near future.
\end{abstract}

\newpage
\section{Introduction}
Several charmless two body decay modes of $B_{u,d}$ mesons have been observed
at CLEO\cite{1,2}.
These data have
provided
interesting information about the Standard Model (SM)\cite{3,4,5,6,7}.
With increased luminosities for B-factories at CLEO,
KEK and SLAC, more
useful information about charmless $B_{u,d}$ decays will be obtained.
The SM can then be tested in more details.
At present only
some weak upper limits on the branching ratios have been obtained\cite{8} for
charmless $B_s$ decays.
However, more data on $B_s$ decays will become available from
hadron colliders, such as CDF, D0, HERAb, BTeV and LHCb in the future.
These data
will help to further test the SM\cite{9}.
Theoretical predictions, on the other hand, are limited by
our inability to reliably calculate hadronic matrix elements related to
$B$ decays although progresses have been made in recent years \cite{10}.
In the lack of reliable calculations, attempts have
been made to extract useful information from symmetry considerations.
$SU(3)$ flavor symmetry\cite{11,12,13} is one of the symmetries which has attracted a lot of
attentions recently.
It has been shown that using SU(3) symmetry it is possible to
constrain\cite{7,14} and to
determine\cite{6}
one of the fundamental parameters $\gamma$ in the SM
for CP violation by measuring several charmless hadronic
B meson decay modes. SU(3) flavor symmetry also predicts many interesting
relations between
CP violating observables in the SM, such as rate differences between $\Delta S = 0$ and
$\Delta S = -1$ B decays. These relations will provide important test for the SM.

CP violating rate differences $\Delta(B\to a_1 a_2) = \Gamma(\bar B \to a_1 a_2)
-\Gamma(B\to \bar a_1
\bar a_2)$ for B decays have been studied before\cite{15,16,17}.
Here $a_i$ can be  one of the octet pseudo-scalars $P$
or one of the octet vector
mesons $V$, respectively.
A complete
study for the charmless $B\to P P$ case was carried out in Ref.\cite{16}.
For $B\to VV$,
as long as rate differences are concerned,
the situation is similar to $B\to PP$.
For $B\to PV$, the situation is more complicated because
there are more independent SU(3) invariant decay amplitudes and more relations exist for these
decays. In this paper we will concentrate on CP violating rate difference 
relations for $B\to P V$ decays in
the SM using SU(3) flavor symmetry.
$B \to PV$ decays using SU(3) flavor symmetry have also been studied in
the literature recently in Ref. \cite{16a,16b} 
with different emphases. Keeping
the leading order contributions, Ref. \cite{16a} 
studied available data from CLEO\cite{1,2}
and obtained information about the phase angle $\gamma$ and proposed new
tests for the SM. Keeping all contributions Ref.\cite{16b} studied electroweak
penguin effects and information about $\gamma$. In our analysis we also
keep all contributions, but with the emphasis on possible relations between
rate differences which have not been studied before in the literature for
B to PV decays.                   
At present CLEO collaboration has measured several $B\to PV$ modes. In the near future,
B factories will measure many more of these decays with more precision 
and rate differences may be measured.
CP violation in the SM will
be tested using $B \to P V$ decays.

The paper is arranged as following. In section II we present the SU(3) decay amplitudes for
charmless $B\to PV$ decays. In section III we study relations between rate
differences for various B to PV
decays. In section IV, we estimate SU(3) breaking effects on the
relations obtained. And in section V we draw our conclusions.

\section{$SU(3)$ decay amplitudes for charmless $B\to PV$}

The quark level effective Hamiltonian up to one loop level in
electroweak interaction
for charmless hadronic $B$ decays,
including the QCD corrections to the matrix elements, can be written as
\begin{eqnarray}
 H_{eff}^q = {G_{F} \over \sqrt{2}} [V_{ub}V^{*}_{uq} (c_1 O_1 +c_2 O_2)
   - \sum_{i=3}^{12}(V_{ub}V^{*}_{uq}c_{i}^{uc} +V_{tb}V_{tq}^*
   c_i^{tc})O_{i}].
\end{eqnarray}
The coefficients
$c_{1,2}$ and $c_i^{jk}=c_i^j-c_i^k$, with $j$ indicates the internal quark,
are the Wilson Coefficients (WC). These
WC's have been evaluated by several groups\cite{18},
with $|c_{1,2}|>> |c_i^j|$.
In the above the factor $V_{cb}V_{cq}^*$ has
been eliminated using the unitarity property of the KM matrix. The
operators $O_i$ are defined as\cite{18},

\begin{eqnarray}
\begin{array}{ll}
O_1=(\bar q_i u_j)_{V-A}(\bar u_i b_j)_{V-A}\;, &
O_2=(\bar q u)_{V-A}(\bar u b)_{V-A}\;,\\
O_{3,5}=(\bar q b)_{V-A} \sum _{q'} (\bar q' q')_{V \mp A}\;,&
O_{4,6}=(\bar q_i b_j)_{V-A} \sum _{q'} (\bar q'_j q'_i)_{V \mp A}\;,\\
O_{7,9}={ 3 \over 2} (\bar q b)_{V-A} \sum _{q'} e_{q'} (\bar q' q')_{V \pm A}\;,\hspace{0.3in} &
O_{8,10}={ 3 \over 2} (\bar q_i b_j)_{V-A} \sum _{q'} e_{q'} (\bar q'_j q'_i)_{V \pm A}\;,\\
O_{11}={g_s\over 16\pi^2}\bar q \sigma_{\mu\nu} G^{\mu\nu} (1+\gamma_5)b\;,&
O_{12}={Q_b e\over 16\pi^2}\bar q \sigma_{\mu\nu} F^{\mu\nu} (1+\gamma_5)b.
\end{array}
\end{eqnarray}
where $(\bar a b)_{V-A} = \bar a \gamma_\mu (1-\gamma_5) b$, $G^{\mu\nu}$ and
$F^{\mu\nu}$ are the field strengths of the gluon and photon, respectively.

At the hadronic level, the decay amplitude for $B\to P V$ can be generically written as

\begin{eqnarray}
A(B\to P V) = <P\;V|H_{eff}^q|B> = V_{ub}V^*_{uq} T(q) + V_{tb}V^*_{tq}P(q)\;,
\label{aa}
\end{eqnarray}
where $T(q)$ contains contributions from
the $tree$ operators $O_{1,2}$ as well as $penguin$ operators $O_{3-12}$
due to charm and up
quark loop corrections to the matrix elements,
while $P(q)$ contains contributions purely from
$penguin$ due to top and charm quarks in loops.
We would like to clarify the notation used here. The amplitude $T$
in eq. \ref{aa} is usually called the ``tree'' amplitude which will also
be referred to later on in the paper. One should, however,
 keep in mind that
it contains the usual tree
current-current contributions proportional to $c_{1,2}$ and also the
u and c penguin contributions proportional to $c_i^u-c_i^c$ with
$i=3-12$ which is small due to cancellation between contributions from up and charm
 quarks in loops.
Also, in general, it contains long distance contributions corresponding
to internal u and c generated intermediate hadron states\cite{19}.
In our later analysis, we do not distinguish between the tree and the penguin
contributions in the amplitude $T$ unless specifically indicated.

The relative strength of the amplitudes
$T$ and $P$ is predominantly determined by their corresponding WC's in the
effective Hamiltonian.
For $\Delta S = 0$ charmless decays, the dominant contributions are due to the
tree operators $O_{1,2}$ and the penguin operators are suppressed by smaller
WC's. Whereas for $\Delta S =-1$ decays, because the penguin contributions are
enhanced by a factor of $V_{tb} V_{ts}^*/V_{ub}V_{us}^*\approx 55$\cite{8}
compared with the
tree contributions, penguin effects dominate the
decay amplitudes. In this case the electroweak penguins
can also play a very important role\cite{20}.
In all $\Delta S =0$ and $\Delta S =-1$ processes, both the tree and penguin
contributions have to be present in order to have rate differences.
One should carefully keep track of all different
contributions.

The operators $O_{1,2}$, $O_{3-6, 11}$, and $O_{7-10,12}$ transform under SU(3)
symmetry as $\bar 3_a + \bar 3_b +6 + \overline {15}$,
$\bar 3$, and $\bar 3_a + \bar 3_b +6 + \overline {15}$, respectively.
These properties enable us to
write the decay amplitudes for $B\to P V$ in only a few SU(3) invariant
amplitudes.

For the $T(q)$ amplitude, for example, we have\cite{12}
\begin{eqnarray}
T(q)&=& A_{\bar 3}(T)B_i H(\bar 3)^i (V_l^k M_k^l)\nonumber\\
 &+& C^V_{\bar 3}(T)B_i V^i_kM^k_jH(\bar 3)^j
+ C^M_{\bar 3}(T)B_i M^i_kV^k_jH(\bar 3)^j \nonumber\\
&+& A^V_{6}(T)B_i H(6)^{ij}_k V^l_jM^k_l
+A^M_{6}(T)B_i H(6)^{ij}_k M^l_jV^k_l \nonumber\\
 &+& C^V_{6}(T)B_iV^i_jH(6)^{jk}_lM^l_k
+ C^M_{6}(T)B_iM^i_jH(6)^{jk}_lV^l_k\nonumber\\
&+&A^V_{\overline {15}}(T)B_i H(\overline {15})^{ij}_k V^l_jM^k_l
+A^M_{\overline {15}}(T)B_i H(\overline {15})^{ij}_k M^l_jV^k_l\nonumber\\
 &+&C^V_{\overline{15}}(T)B_iV^i_jH(\overline {15} )^{jk}_lM^l_k
+C^M_{\overline{15}}(T)B_iM^i_jH(\overline {15} )^{jk}_lV^l_k\;,
\label{am}
\end{eqnarray}
where $B_i = (B_u,  B_d,  B_s) = (B^-, \bar B^0, \bar B^0_s)$
is a SU(3) triplet, $M_{i}^j$ and $V_i^j$ are the SU(3) pseudo-scalar and
vector meson
octets, respectively. In our analysis, we will also include the SU(3)
singlets for both the psuedo-scalar and vector mesons, such that $M$ and $V$
become the U(3) nonet, to have some idea
about the decay amplitudes for B decays involving these particles.
The SU(3) invariant amplitudes
$A_i(T)$ and $C_i(T)$ are in general complex due to final state interactions.
The matrices $H(i)$
contain information about
the transformation properties of the operators $O_{1-12}$.

For $q=d$, the non-zero entries of the matrices $H(i)$ are given by\cite{12}
\begin{eqnarray}
H(\bar 3)^2 &=& 1\;,\;\;
H(6)^{12}_1 = H(6)^{23}_3 = 1\;,\;\;H(6)^{21}_1 = H(6)^{32}_3 =
-1\;,\nonumber\\
H(\overline {15} )^{12}_1 &=& H(\overline {15} )^{21}_1 = 3\;,\; H(\overline
{15} )^{22}_2 =
-2\;,\;
H(\overline {15} )^{32}_3 = H(\overline {15} )^{23}_3 = -1\;.
\end{eqnarray}
And for $q = s$, the non-zero entries are\cite{15}
\begin{eqnarray}
H(\bar 3)^3 &=& 1\;,\;\;
H(6)^{13}_1 = H(6)^{32}_2 = 1\;,\;\;H(6)^{31}_1 = H(6)^{23}_2 =
-1\;,\nonumber\\
H(\overline {15} )^{13}_1 &=& H(\overline {15} ) ^{31}_1 = 3\;,\; H(\overline
{15} )^{33}_3 =
-2\;,\;
H(\overline {15} )^{32}_2 = H(\overline {15} )^{23}_2 = -1\;.
\end{eqnarray}
There are similar amplitudes for the penguins. We will indicate these SU(3) invariant amplitudes
by $A_i(P)$ and $C_i(P)$.

The decay amplitudes can be written in terms of the
SU(3) invariant amplitudes as

\begin{eqnarray}
A(\bar B \to P V) &=&
V_{ub}V_{uq}^* (a_i^V A^V_i(T) + a^M_i A_i^M(T) + b_i^V C_i^V(T) +b_i^M C_i^M(T))\nonumber\\
&+&V_{tb}V_{tq}^* (a_i^V A^V_i(P) + a^M_i A_i^M(P) + b_i^V C_i^V(P) +b_i^M C_i^M(P)).
\end{eqnarray}
The index $i$
 is summed over $i = \bar 3, 6, \overline {15}$. For $i=\bar 3$, there is only one
$A_{\bar 3}$ amplitude. We use the convention: $A_{\bar 3}^V = A_{\bar 3}$, $a_{\bar 3}^V = a_{\bar 3}$
and $a_{\bar 3}^MA_{\bar 3}^M=0$.

One can easily obtain the decay amplitudes for $B\to PP$ and $B\to VV$ from the above by
replacing $V^i_j$ by $M^i_j$, and $M^i_j$ by $V^i_j$ in the above expression, respectively.
The SU(3) invariant amplitudes are then replaced by $A_i = A_i^V +A_i^M$ and $C_i = C_i^V+C_i^M$.
In these two cases, due to the anti-symmetric nature in exchanging the upper two indices of
$H^{ij}_k(6)$ and the symmetric structure of the two mesons in the final states,
$C_6-A_6$ always appear together\cite{12} if the singlets in the final states
are removed. Therefore in these cases
there are in total 5 independent SU(3) invariant
amplitudes for each case. However if singlets are included, decay modes
involving singlets in the final states can be used to separate $A_6$ and $C_6$.
In this case there are in total 6 independent amplitudes.
For $B\to P V$, even
the symmetric structure of the two mesons in the final
states is lost, there are relations between
$A_6^i$ and $C_6^i$ if the singlets $\eta_1$ and the combination of
$(\sqrt{2/3})\omega+\phi/\sqrt{3}$ are absent, there are total 10 independent
invariant amplitudes which agree with the analysis in Ref. \cite{16b}.
When the singlets are introduced,
$A_i$ and $C_i$ are all independent. There are total
11 of them. The analysis in this case is more complicated compared with the cases for
$B\to PP$ and $B\to VV$.
Expanding eq. \ref{am}, we obtain the coefficients for each individual decays.
In tables 1 - 3 we list the coefficients $a_i$ and $b_i$ for $\Delta S=0$ $B\to PV$ decays,
and in Tables 4-6 for $\Delta S = -1$ $B\to PV$ decays.

Two remarks are in order: a)
The amplitudes $A_i$ correspond to annihilation contributions, as can be seen from eq.
\ref{am} where $B_i$ is contracted with one of the index in $H(j)$, are small compared with
the amplitudes $C_i$ from model calculations. The smallness of these annihilation amplitudes
can be tested using $B_d \to K^- K^{*+}, K^+ K^{*-}$ and $B_s \to \pi^+ \rho^-, \pi^-\rho^+,
\pi^0 \rho^0, \pi^0 \omega$
because these decays have only annihilation contributions.
b) Many analyses have been carried out using SU(3) classification of quark level
diagrams\cite{13} in the literature. In most cases such  analyses give the same results as
the use of SU(3) invariant amplitudes discussed here. However, in some cases the
classification according to quark level diagrams without care
could be misleading. One should be careful to include all possible contributions
to have a complete study\cite{16b}.
For example, neglecting annihilation contributions,
the operators $O_{1,2}$ would have vanishing contributions to
$B_u \to K^- K^{*0}, \bar K^0 K^{*-}, \pi^- \bar K^{*0}, \bar K^0 \rho^-$,
$B_d \to K^0 \bar K^{*0}, \bar K^0 K^{*0}$, and $B_s \to K^0 \bar K^{*0},
\bar K^0 K^{*0}$.
However, from the Tables, we find that these decays are proportional to
$C_{\bar 3}^{V,M} - C_6^{V,M} - C_{\overline {15}}^{V,M}$. These combinations are not necessarily
zero without additional assumptions. Whether naive diagram analysis is valid has to be tested
experimentally. This can be
achieved by measuring the decay modes mentioned in the remarks a) and b). Without evidence from
experimental data, we have to keep the most general form and carry out analyses accordingly.

\section{CP Violating Rate Differences for Charmless $B\to PV$}

One can find many relations between different $B$ 
decays using SU(3) flavor symmetry to test the SM
in different ways. A
particularly interesting class of relations for $B\to PV$ is the CP violating rate difference
$\Delta(B\to PV) = \Gamma(B\to P V) - \Gamma(\bar B \to \bar P \bar V)$.
Theoretical calculations of rate differences are difficult because one not only needs to
calculate the short distance decay amplitudes but also the long distance contributions,
especially the final state interaction phases.
This is a difficult task at present.
Due to this
reason, even though CP violation is measured for  certain $B$ decay modes and 
compared with hadronic model calculations, one is not sure
if the results agree with the SM predictions.
In this section we will derive several relations which do not depend on the dynamics of
hadronization processes using the SU(3) decay amplitudes
obtained in the previous section. These relations can provide hadronization model independent
tests for CP violation in the SM.

Using the SU(3) decay amplitudes obtained in the previous section, one can find that
some decay amplitudes for $\Delta S = 0$ and $\Delta S = -1$,
$B\to PV$ decays have the following peculiar
form\cite{15,16,17,21} ,

\begin{eqnarray}
A(d) = V_{ub}V_{ud}^* T + V_{tb}V_{td}^* P,\nonumber\\
A(s) = V_{ub}V_{us}^* T + V_{tb}V_{ts}^* P.
\end{eqnarray}

Due to different KM matrix elements involved in $A(d)$ and $A(s)$,
although the amplitudes have some similarities, the branching ratios are
not simply related. However, when considering rate difference,
the situation is dramatically different. Because a simple property
of the KM matrix element\cite{22}, $Im (V_{ub}V_{ud}^*V_{tb}^*V_{td})
=-Im(V_{ub}V_{us}^*V_{tb}^*V_{ts})$, in the SU(3) limit we have

\begin{eqnarray}
\Delta(d) =- \Delta(s),
\end{eqnarray}
where $\Delta(i) = (|A(i)|^2-|\bar A(i)|^2) \lambda_{ab}/(8\pi m_B)$ is the
CP violating rate difference defined earlier and
$\lambda_{ab} = \sqrt{1-2(m_a^2+m_b^2)/m_B^2 + (m_a^2-m_b^2)^2/m_B^4}$ with
$m_{a,b}$ being the masses of the two particles in the final state.

The above non-trivial equality does not
depend on the numerical values of the final state re-scattering phases.
Of course these relations are true only for models
with three generations.
Therefore they provide dynamic model independent tests for the three generation
Standard Model.

We find the following equalities:

\begin{eqnarray}
(1)&\;\;& \Delta(B_u \to K^- K^{*0}) = - \Delta (B_u \to \pi^- \bar K^{*0})\;,\nonumber\\
(2)&& \Delta(B_d \to {\bar K}^0 K^{*0}) = - \Delta (B_s \to K^0 {\bar K}^{*0})\;,\nonumber\\
(3)&& \Delta(B_u \to  K^0 K^{*-}) = - \Delta (B_u \to \bar K^0  \rho^-)\;,\nonumber\\
(4)&& \Delta(B_d \to K^0 {\bar K}^{*0}) = - \Delta (B_s \to {\bar K}^0 K^{*0})\;,\nonumber\\
(5)&& \Delta(B_d \to \pi^- \rho^+) = - \Delta (B_s \to K^- K^{*+})\;,\nonumber\\
(6)&& \Delta (B_s\to\pi^-K^{*+}) = -\Delta(B_d \to  K^-\rho^+) \;,\nonumber\\
(7)&& \Delta(B_d \to \pi^+ \rho^-) = - \Delta (B_s \to K^+ K^{*-})\;,\nonumber\\
(8)&& \Delta (B_s\to K^+\rho^-) = -\Delta(B_d\to \pi^+ K^{*-})\;,\nonumber\\
(9)&& \Delta(B_u \to \eta_1 \rho^-) = - \Delta (B_u \to \eta_1 K^{*-})\;,\nonumber\\
(10)&&  \Delta (B_s \to \eta_1 K^{*0})= -\Delta(B_d \to \eta_1 {\bar K}^{*0}) \;,\nonumber\\
(11)&& \Delta(B_d \to K^- K^{*+}) = - \Delta (B_s \to \pi^- \rho^+)\;,\nonumber\\
(12)&& \Delta(B_d \to K^+ K^{*-}) = - \Delta (B_s \to \pi^+ \rho^-).
\end{eqnarray}

If it turns out that the annihilation contributions are all small, as can be
tested in
$B_d \to K^- K^{*+}$, $K^+ K^{*-}$ and $B_s \to \pi^+ \rho^-$, $\pi^- \rho^+$,
$\pi^0\rho^0$, $\pi^0 \omega$,
there are additional relations for rate differences.
We find the following equalities,

\begin{eqnarray}
\begin{array}{ll}
(1)=(2)\;,\hspace{0.6in}
&(3)=(4)\;,\\
(5)=(6)\;,
&(7)=(8).
\end{array}
\end{eqnarray}

\section{SU(3) Breaking Effects}

The relations obtained in the previous section
hold in the SU(3) limit.
The SU(3) symmetry need to be tested experimentally.
To this end we comment that
SU(3) predictions for $B\to PV$ with charmed vector meson V can be independently tested by
using $B\to D^* \pi$ and $B\to D^* K$. These
decays receive contributions from the tree operators $O_{1,2}$ only. This provides a clear test for
SU(3) symmetry free from penguin contaminations.
In the SU(3) symmetry, the ratio of these branching ratios is equal to
$r=Br(B\to D^* \pi)/ Br(B\to D^* K) = |V_{ud}/V_{us}|^2$. Factorization calculation gives a
SU(3) breaking factor and $r=(f_\pi^2/f_k^2)|V_{ud}/V_{us}|^2$.
This can be tested experimentally. Results
obtained from this will give us a good guidance on how well CP violation can be tested in
charmless $B \to P V$ decays.

If SU(3) is broken these relations need to be modified.
We now study how
these relations are modified when SU(3) breaking effects are included.
Since no reliable calculational tool exists,
in the following we will use factorization
approximation neglecting the annihilation contributions
to estimate the SU(3) breaking effects.
We find\cite{3,9}

\begin{eqnarray}
(1)&\;\;& \Delta(B_u \to K^- K^{*0}) =
 -({{F^{B \to K}_1(m^2_{K^*})} \over {F^{B \to \pi}_1(m^2_{K^*})}})^2
\Delta (B_u \to \pi^- \bar K^{*0})\;,\nonumber\\
(2)&& \Delta(B_d \to {\bar K}^0 K^{*0}) =
 - ({{F^{B \to K}_1(m^2_{K^*}} \over {F^{B_s \to K}_1(m^2_{K^*})}})^2
\Delta (B_s \to K^0 {\bar K}^{*0})\;,\nonumber\\
(3)&& \Delta(B_u \to  K^0 K^{*-}) =
- ({{A^{B \to K^*}_0(m^2_K)} \over {A^{B \to \rho}_0(m^2_\rho)}})^2
\Delta (B_u \to \bar K^0  \rho^-)\;,\nonumber\\
(4)&& \Delta(B_d \to K^0 {\bar K}^{*0}) =
 - ({{A^{B \to K^*}_0(m^2_K)} \over {A^{B_s \to K^*}_0(m^2_K)}})^2
\Delta (B_s \to {\bar K}^0 K^{*0})\;,\nonumber\\
(5)&& \Delta(B_d \to \pi^- \rho^+) =
 - ({ {f_{\pi} A^{B \to \rho}_0(m^2_\pi)} \over {f_K A^{B_s \to K^*}_0(m^2_K)}})^2
\Delta (B_s \to K^- K^{*+})\;,\nonumber\\
(6)&& \Delta (B_s\to\pi^-K^{*+}) =
 - ({ {f_{\pi} A^{B_s \to K^*}_0(m^2_\pi)} \over {f_K A^{B \to \rho}_0(m^2_K)}})^2
\Delta(B_d \to  K^-\rho^+) \;,\nonumber\\
(7)&& \Delta(B_d \to \pi^+ \rho^-) =
 - ({ {f_{\rho} F^{B \to \pi}_1(m^2_\rho)} \over {f_{K^*} F^{B_s \to K}_1(m^2_{K^*}}})^2
\Delta (B_s \to K^+ K^{*-})\;,\nonumber\\
(8)&& \Delta (B_s\to K^+\rho^-) = - ({ {f_{\rho} F^{B_s \to K}_1(m^2_\rho)} \over {f_{K^*} F^{B \to \pi}_1(m^2_{K^*})}})^2
\Delta(B_d\to \pi^+ K^{*-}).
\end{eqnarray}
In the above we have not listed the corrections for the equalities (9), (10), (11) and (12).
The corrections to the decay modes in (9) and (10) involving
$\eta_1$ are complicated
because there are two corrections to the amplitudes and also because
mixings between $\eta_1$ and $\eta_8$. It is difficult to use these decay modes to have
clear tests. The decay modes in (11) and (12) are all pure annihilation
type of decays which are zero in the naive factorization approximation and likely to be small. We will not discuss these decay modes further here.

The decay modes in (1), (2), (3) and (4) do not have contributions
from $O_{1,2}$
in the factorization approximation.
Therefore their rate differences are expected
to be very small. However one should be careful in drawing this conclusion, as
already mentioned earlier, due to the fact
that the vanishing contributions from $O_{1,2}$ to these
modes is not a SU(3) prediction even when annihilation contributions are neglected.
It may happen that the rate differences are larger than expected. We have to
wait experimental data to tell us more.

At $e^+ e^-$ $B$ factories the best chances to test the above listed relations are
from (5), (6), (7) and (8). To a good approximation, we have

\begin{eqnarray}
(a)&\;\;\;\;&\Delta (B_d \to \pi^- \rho^+) \approx - ({f_\pi A^{B\to \rho}_0(m^2_\pi)
\over f_K A^{B\to \rho}_0(m^2_K)})^2 \Delta (B_d \to K^- \rho^+)\;,\nonumber\\
(b)&&\Delta (B_d \to \pi^+ \rho^-) \approx
-({f_\rho F_1^{B\to \pi}(m^2_\rho)\over f_{K^*} F_1^{B\to \pi}(m^2_{K^*})})^2
\Delta (B_d \to \pi^+ K^{*-})\;.
\end{eqnarray}
The ratios $A^{B\to \rho}(m^2_\pi)/A^{B\to\rho}_0(m^2_K)$ and
$F^{B\to \pi}_1(m^2_\rho)/F^{B\to \pi}_1(m^2_{K^*})$ are all close to one[3]. The
corrections are dominated by the decay constants $f_i$. The relation (a) has,
then, a larger correction than (b) because $f_\rho/f_{K^*} \approx 1$ and
$f_\pi/f_K \approx 0.82$.

The decay branching ratios for $B_d \to \pi^+ K^{*-}$ and the combined
$B_d \to \pi^+\rho^-,\;\pi^-\rho^+$ have been measured at CLEO\cite{2}  with
$Br(B_d \to \pi^+ K^{*-}) = (22^{+8+4}_{-6-5})\times 10^{-6}$, and
$Br(B_d \to \pi^+\rho^-+\pi^-\rho^+) = (35^{+11}_{-10}\pm 5)\times 10^{-6}$,
respectively. There is only an upper bound for $Br(B_d \to K^-\rho^+)
<25\times 10^{-6}$ at the 90\% c.l..
If we normalize the rate differences by a common branching ratio,
for example, $Br(B_d \to \pi^+ K^{*-})$, the
normalized rate differences for $B_d \to \pi^-\rho^+$, $K^-\rho^+$ are of order
a few percent\cite{3}, while for $B_d \to \pi^+ \rho^-$, $\pi^+ K^{*-}$ are of order
20\%\cite{3}. These can be measured at B factories with good precision.
The relations predicted can be tested. Even earlier a test can be performed with
the combined data

\begin{eqnarray}
\Delta (B_d \to \pi^-\rho^+) + \Delta (B_d\to \pi^+\rho^-)
\approx - ({f_\pi\over f_K})^2 \Delta (B_d \to K^-\rho^+)
- ({f_\rho\over f_{K^*}})^2 \Delta (B_d \to \pi^+ K^{*-})\;.\nonumber
\end{eqnarray}

When $B_s$ decay modes are measured, such as at CDF, D0, BTeV, HERAb, and LHCb
all modes in (5), (6), (7) and (8) can provide good tests for the Standard Model.

\section{Conclusions and Discussions}

We would like to point out that the relations obtained for
the CP violating rate differences also hold for any three generation
models in which flavor changing and CP violating
source is solely due to KM matrix,
such as multi-Higgs doublet models with neutral flavor
current conservation by exchanging of Higgs bosons at tree level and model with anomalous three gauge boson
couplings as new physics source.

To conclude we find that
although the relations between branching ratios of $\Delta S = 0$
and $\Delta S = -1$ processes are complicated, there are simple
relations between some of the $\Delta S =0$ and $\Delta S = -1$
rate differences
 due to the unitarity property of the KM matrix,
such as $ \Delta(B\to \pi^+ \rho^-) = -\Delta(B\to \pi^+ K^{*-})$,
$\Delta(B\to \pi^- \rho^+) = - \Delta(K^- \rho^+)$. We also estimated
SU(3) breaking effects
using factorization approximation.
The relations involving $B_d$ and $B_u$ decays can soon be tested by
CLEO, Babar and Belle collaborations. While the relations
involving $B_s$ decays 
can be carried out at hadron colliders, such as CDF, D0, BTeV, HERAb and LHCb.
CP violation can be tested in a model independent way 
for the Standard Model using the relations derived in this paper soon.
We emphasis that the importance of these tests is their independence from
hadronic models.
We urge our experimental colleagues to carry out such analyses.

\acknowledgements This work was partially supported by DOE of USA
under grant number DE-FG06-85ER40224 and by NSC of R.O.C under
grant number NSC 89-2112-M-002-016.

\newpage
\noindent
Table 1. SU(3) decay amplitudes for $\Delta S = 0$, $B_u\to PV$ decays.
\vspace{0.5cm}

\begin{tabular}{|l|rrrrrrrrrrrrr|}\hline
Decay Mode && $a_{\bar 3}$ & $a^V_6$ & $a^M_6$
& $a^V_{\overline {15}}$ & $a^M_{\overline {15}}$
& $b^V_{\bar 3}$ & $b^M_{\bar 3}$ & $b^V_6$ & $b^M_6$
& $b^V_{\overline {15}}$ & $b^M_{\overline {15}}$& \\ \hline\hline
$B_u\to K^-K^{*0}$& (&0&1&0&3&0&0&1&0&-1&0&-1&)        \\
$B_u\to K^0 K^{*-} $& (&0&0&1&0&3&1&0&-1&0&-1&0&)         \\
$B_u\to\eta_8 \rho ^- $&$1 \over \sqrt{6}$ (&0&1&1&3&3&1&1&-3&1&3&3&)      \\
$B_u\to\eta_1\rho ^-$&$1 \over \sqrt{3}$ (&0&1&1&3&3&1&1&0&1&0&3&)     \\
$B_u\to\pi^0 \rho ^-$&$1 \over \sqrt{2}$ (&0&1&-1&3&-3&-1&1&-1&1&5&3&)  \\
$B_u\to\pi^- \phi$ &(&0&0&0&0&0&0&0&0&1&0&-1&)  \\
$B_u\to\pi^- \rho^0$&$1 \over \sqrt{2 }$ (&0&-1&1&-3&3&1&-1&1&-1&3&5&)  \\
$B_u\to\pi^- \omega$&$1 \over\sqrt{2 }$ (&0&1&1&3&3&1&1&1&-1&3&1&)   \\ \hline
\end{tabular}

\newpage
\noindent
Table 2. SU(3) decay amplitudes for $\Delta S = 0$, $B_d \to PV$ decays.
\vspace{0.5cm}

\begin{tabular}{|l|rrrrrrrrrrrrr|}\hline
Decay Mode && $a_{\bar 3}$ & $a^V_6$ & $a^M_6$
& $a^V_{\overline {15}}$ & $a^M_{\overline {15}}$
& $b^V_{\bar 3}$ & $b^M_{\bar 3}$ & $b^V_6$ & $b^M_6$
& $b^V_{\overline {15}}$ & $b^M_{\overline {15}}$& \\ \hline\hline
$B_d\to K^0 {\bar K}^{*0}$       &(&1&1&0&-1&-2&1&0&-1&0&-1&0&)   \\
$B_d\to\bar K^0 K^{*0}$      &(&1&0&1&-2&-1&0&1&0&-1&0&-1&)    \\
$B_d\to K^- K^{*+}$      &(&1&-1&1&3&-1&0&0&0&0&0&0& )   \\
$B_d\to\pi^- \rho^+$   &(&1&-1&0&3&-2&1&0&1&0&3&0&)    \\
$B_d\to\pi^+ \rho^-$      &(&1&0&-1&-2&3&0&1&0&1&0&3&)    \\
$B_d\to K^+ K^{*-}$     &(&1&1&-1&-1&3&0&0&0&0&0&0& )   \\
$B_d\to\eta_8 \phi$&$1\over \sqrt{6 }$ (&-2&-2&-2&2&2&0&0&0&1&0&-1& )   \\
$B_d\to\eta_8 \omega$&$1 \over{2 \sqrt{3 }}$ (&2&-1&-1&1&1&1&1&-3&-1&3&1&)    \\
$B_d\to\eta_8 \rho^0$&$1 \over{2 \sqrt{3 }}$ (&0&-1&-1&5&5&-1&-1&3&-1&-3&5&)    \\
$B_d\to\eta_1 \phi$&$1 \over \sqrt{3 }$  (&1&1&1&-1&-1&0&0&0&1&0&-1& )   \\
$B_d\to\eta_1 \omega$&$1 \over \sqrt{6 }$      (&2&-1&-1&1&1&1&1&0&-1&0&1&)    \\
$B_d\to\eta_1 \rho^0$&$1 \over \sqrt{6 }$      (&0&-1&-1&5&5&-1&-1&0&-1&0&5&)    \\
$B_d\to\pi^0 \rho^0$&$1 \over 2$      (&2&-1&-1&1&1&1&1&1&1&-5&-5&)    \\
$B_d\to\pi^0 \phi$&$1 \over \sqrt{2 }$      (&0&0&0&0&0&0&0&0&-1&0&1&)    \\
$B_d\to\pi^0 \omega$&$1 \over 2$ (&0&-1&-1&5&5&-1&-1&-1&1&5&-1&)    \\
 \hline
\end{tabular}

\newpage
\noindent
Table 3. SU(3) decay amplitudes for $\Delta S = 0$, $B_s \to PV$ decays.
\vspace{0.5cm}

\begin{tabular}{|l|rrrrrrrrrrrrr|}\hline
Decay Mode && $a_{\bar 3}$ & $a^V_6$ & $a^M_6$
& $a^V_{\overline {15}}$ & $a^M_{\overline {15}}$
& $b^V_{\bar 3}$ & $b^M_{\bar 3}$ & $b^V_6$ & $b^M_6$
& $b^V_{\overline {15}}$ & $b^M_{\overline {15}}$& \\ \hline\hline
$B_s\to K^0 \phi$&(&0&0&-1&0&-1&1&0&-1&1&-1&-1&)    \\
$B_s\to K^0 \rho^0$&$1 \over \sqrt{2 }$ (&0&1&0&1&0&0&-1&0&-1&0&5& )   \\
$B_s\to K^0 \omega$&$1 \over \sqrt{2 }$      (&0&-1&0&-1&0&0&1&0&-1&0&1& )   \\
$B_s\to K^+ \rho^-$&      (&0&-1&0&-1&0&0&1&0&1&0&3&)    \\
$B_s\to\pi^- K^{*+}$&      (&0&0&-1&0&-1&1&0&1&0&3&0 & )  \\
$B_s\to\pi^0 K^{*0}$&$1 \over \sqrt{2 }$      (&0&0&1&0&1&-1&0&-1&0&5&0& )   \\
$B_s\to\eta_1 K^{*0}$&$1 \over \sqrt{3 }$      (&0&-1&-1&-1&-1&1&1&0&-1&0&-1& )   \\
$B_s\to\eta_8 K^{*0}$&$1 \over \sqrt{6 }$      (&0&2&-1&2&-1&1&-2&-3&2&3&2 & )  \\
 \hline
\end{tabular}

\newpage
\noindent
Table 4. SU(3) decay amplitudes for $\Delta S = -1$, $B_u\to PV$ decays.
\vspace{0.5cm}

\begin{tabular}{|l|rrrrrrrrrrrrr|}\hline
Decay Mode && $a_{\bar 3}$ & $a^V_6$ & $a^M_6$
& $a^V_{\overline {15}}$ & $a^M_{\overline {15}}$
& $b^V_{\bar 3}$ & $b^M_{\bar 3}$ & $b^V_6$ & $b^M_6$
& $b^V_{\overline {15}}$ & $b^M_{\overline {15}}$& \\ \hline\hline
$B_u\to\pi^- {\bar K}^{*0}$&(&0&1&0&3&0&0&1&0&-1&0&-1 &)   \\
$B_u\to\eta_8 K^{*-}$&$1 \over \sqrt{6 }$      (&0&1&-2&3&-6&-2&1&0&1&6&3&)    \\
$B_u\to\eta_1 K^{*-}$&$1 \over \sqrt{3 }$      (&0&1&1&3&3&1&1&0&1&0&3& )   \\
$B_u\to K^-  \phi$&      (&0&1&0&3&0&0&1&0&0&0&-2& )   \\
$B_u\to K^- \rho^0$&$1 \over \sqrt{2 }$      (&0&0&1&0&3&1&0&1&-2&3&4&)    \\
$B_u\to K^- \omega$&$1 \over \sqrt{2 }$      (&0&0&1&0&3&1&0&1&0&3&2& )   \\
$B_u\to\bar K^0 \rho^-$&$1 $      (&0&0&1&0&3&1&0&-1&0&-1&0&)    \\
$B_u\to\pi^0 K^{*-}$&$1 \over \sqrt{2 }$      (&0&1&0&3&0&0&1&-2&1&4&3&  )  \\
 \hline
\end{tabular}
\vspace{1cm}

\noindent
Table 5. SU(3) decay amplitudes for $\Delta S =-1$, $B_d \to PV$ decays.
\vspace{0.5cm}

\begin{tabular}{|l|rrrrrrrrrrrrr|}\hline
Decay Mode && $a_{\bar 3}$ & $a^V_6$ & $a^M_6$
& $a^V_{\overline {15}}$ & $a^M_{\overline {15}}$
& $b^V_{\bar 3}$ & $b^M_{\bar 3}$ & $b^V_6$ & $b^M_6$
& $b^V_{\overline {15}}$ & $b^M_{\overline {15}}$& \\ \hline\hline
$B_d\to\bar K^0 \rho^0$&$1 \over \sqrt{2 }$      (&0&0&1&0&1&-1&0&1&-2&1&4&)    \\
$B_d\to\bar K^0 \omega$&$1 \over \sqrt{2 }$      (&0&0&-1&0&-1&1&0&-1&0&-1&2&)    \\
$B_d\to\bar K^0 \phi$    &(&0&-1&0&-1&0&0&1&0&0&0&-2&)    \\
$B_d\to\pi^+ {\bar K}^{*-}$      &(&0&-1&0&-1&0&0&1&0&1&0&3& )   \\
$B_d\to K^- \rho^+$     &(&0&0&-1&0&-1&1&0&1&0&3&0&)    \\
$B_d\to\pi^0 {\bar K}^{*0}$&$1 \over \sqrt{2 }$      (&0&1&0&1&0&0&-1&-2&1&4&1& )   \\
$B_d\to\eta_1 {\bar K}^{*0}$&$1 \over \sqrt{3 }$      (&0&-1&-1&-1&-1&1&1&0&-1&0&-1&)    \\
$B_d\to\eta_8 {\bar K}^{*0}$&$1 \over \sqrt{6 }$      (&0&-1&2&-1&2&-2&1&0&-1&6&-1& )   \\
 \hline
\end{tabular}

\newpage
\noindent
Table 6. SU(3) decay amplitudes for $\Delta S = -1$, $B_s \to PV$ decays.
\vspace{0.5cm}

\begin{tabular}{|l|rrrrrrrrrrrrr|}\hline
Decay Mode && $a_{\bar 3}$ & $a^V_6$ & $a^M_6$
& $a^V_{\overline {15}}$ & $a^M_{\overline {15}}$
& $b^V_{\bar 3}$ & $b^M_{\bar 3}$ & $b^V_6$ & $b^M_6$
& $b^V_{\overline {15}}$ & $b^M_{\overline {15}}$& \\ \hline\hline
$B_s\to K^0 {\bar K}^{*0}$&      (&1&0&1&-2&-1&0&1&0&-1&0&-1&)    \\
$B_s\to\bar K^0 K^{*0}$&      (&1&1&0&-1&-2&1&0&-1&0&-1&0&  )  \\
$B_s\to K^- K^{*+}$&     (&1&-1&0&3&-2&1&0&1&0&3&0& )   \\
$B_s\to \pi^- \rho^+$&      (&1&-1&1&3&-1&0&0&0&0&0&0&)    \\
$B_s\to \pi^+ \rho^-$&      (&1&1&-1&-1&3&0&0&0&0&0&0& )   \\
$B_s\to K^+ K^{*-}$&      (&1&0&-1&-2&3&0&1&0&1&0&3& )   \\
$B_s\to \eta_8 \phi$&$\sqrt{ 2 \over 3 }$      (&-1&0&0&2&2&-1&-1&0&0&3&2& )   \\
$B_s\to \eta_8 \rho^0$&$1 \over \sqrt{3 }$      (&0&-1&-1&2&2&0&0&0&2&0&-4& )   \\
$B_s\to\eta_8 \omega$&$1 \over \sqrt{3 }$      (&1&0&0&1&1&0&0&0&0&0&-2& )   \\
$B_s\to\eta_1 \phi$&$1 \over \sqrt{3 }$      (&1&0&0&-2&-2&1&1&0&0&0&-2& )   \\
$B_s\to\eta_1 \omega$&$\sqrt{ 2 \over 3}$      (&1&0&0&1&1&0&0&0&0&0&1&  )  \\
$B_s\to\eta_1 \rho^0$&$\sqrt{ 2 \over 3}$      (&0&-1&-1&2&2&0&0&0&-1&0&2&)    \\
$B_s\to\pi^0 \rho^0$&      (&1&0&0&1&1&0&0&0&0&0&0& )   \\
$B_s\to\pi^0 \phi$&$\sqrt{2 }$      (&0&0&0&0&0&0&0&-1&0&2&0& )   \\
$B_s\to\pi^0 \omega$&     (&0&-1&-1&2&2&0&0&0&0&0&0&  )  \\
\hline
\end{tabular}
\end{document}